\documentclass[a4paper,11pt]{article}
\pdfoutput=1 

\usepackage{jheppub} 

\usepackage[T1]{fontenc} 
\usepackage{empheq,mathtools}
\usepackage{cancel,slashed}

\def\tsp{\hspace{0.083333em}}
\def\arXiv#1{\href{http://arxiv.org/abs/#1}{arXiv:#1}}
\def\arXiv#1#2{\href{http://arxiv.org/abs/#1}{arXiv:#1}}
\def\arXivid#1#2{\href{http://arxiv.org/abs/#1/#2}{#1/#2}}
\def\doi#1#2{\href{http://doi.org/#1}{#2}}

\title{\boldmath Analytic critical points of charged R\'{e}nyi entropies from hyperbolic black holes}

\author{Jie Ren}

\affiliation{School of Physics, Sun Yat-sen University, Guangzhou, 510275, China}

\emailAdd{renjie7@mail.sysu.edu.cn}

\abstract{We analytically study phase transitions of holographic charged R\'{e}nyi entropies in two gravitational systems dual to the $\mathcal{N}=4$ super-Yang-Mills theory at finite density and zero temperature. The first system is the Reissner-Nordstr\"{o}m-AdS$_5$ black hole, which has finite entropy at zero temperature. The second system is a charged dilatonic black hole in AdS$_5$, which has zero entropy at zero temperature. Hyperbolic black holes are employed to calculate the R\'{e}nyi entropies with the entangling surface being a sphere. We perturb each system by a charged scalar field, and look for a zero mode signaling the instability of the extremal hyperbolic black hole. Zero modes as well as the leading order of the full retarded Green's function are analytically solved for both systems, in contrast to previous studies in which only the IR (near horizon) instability was analytically treated.}

\begin{document} 
\maketitle
\flushbottom

\section{Introduction}
\label{sec:intro}
R\'{e}nyi entropies as a generalization of the entanglement entropy play an important role in characterizing quantum systems. Generally, they are difficult to calculate in quantum field theories. The AdS/CFT correspondence provides a powerful tool to study some strongly interacting quantum field theories in the large $N$ limit in terms of classical gravity~\cite{Ryu:2006bv,Ryu:2006ef,Casini:2011kv,Hung:2011nu}. As an example, R\'{e}nyi entropies with the entangling surface being a sphere can be calculated in terms of hyperbolic black holes~\cite{Casini:2011kv,Hung:2011nu}. The parameter $n$ of R\'{e}nyi entropies $S_n$ is related to the temperature of hyperbolic black holes: larger $n$ means a lower temperature. If we perturb the black hole by a scalar field, instability may happen as the temperature decreases. Consequently, the hyperbolic black holes will develop a scalar hair, which implies that R\'{e}nyi entropies will have a phase transition in $n$ \cite{Belin:2013dva}.

Charged R\'{e}nyi entropies was studied in as a generalization of R\'{e}nyi entropies for finite density systems \cite{Belin:2013uta,Belin:2014mva}. Similarly, they can be calculated in terms of charged hyperbolic black holes if the entangling surface is a sphere. As the temperature decreases, instability may happen if a charged scalar field is included in the system~\cite{Belin:2014mva}, which can be regarded as holographic superconductors \cite{Gubser:2008px,Hartnoll:2008vx,Hartnoll:2008kx} in hyperbolic space. Since lower temperature black holes are typically more unstable, examining instabilities at zero temperature will help to understand phase transitions at finite temperature. Therefore, we focus on the onset of instability of zero-temperature hyperbolic charged black holes.

To find the instability, we perturb the system by a charged scalar field, and examine non-analyticities of the retarded Green's function $G(\omega)$. There are two types of instabilities of a scalar field in the hyperbolic Reissner-Nordstr\"{o}m-AdS (RN-AdS) black hole background at zero temperature (they can happen at the same time):
\begin{itemize}
\item[(i)] In the retarded Green's function, when a pole in the lower half complex-$\omega$ plane moves to the upper half plane, there is an instability. A zero mode is defined by a pole at $\omega=0$, and it signals the onset of an instability. To obtain a Green's function near the critical point, we need to match the IR and UV data \cite{Faulkner:2009wj}.

\item[(ii)] The IR (near-horizon) geometry of the extremal black hole is AdS$_2\times\mathbb{H}_{d-1}$.  When the Breitenlohner-Freedman (BF) bound for this AdS$_2$ is violated \cite{Dias:2010ma}, i.e., the scaling exponent for AdS$_2$ becomes imaginary, there is an IR instability. Infinite number of poles will appear at the upper half complex-$\omega$ plane, with the origin being an accumulation point~\cite{Faulkner:2009wj}.
\end{itemize}

The second type of instability was analytically calculated due to the AdS$_2$ factor in the IR geometry \cite{Belin:2013dva,Belin:2014mva}. Neutral hyperbolic black holes only have this type of instability at zero temperature.\footnote{Both neutral and charged hyperbolic black holes have an AdS$_2$ factor at zero temperature. As a comparison, for planar black holes, only the charged one has an AdS$_2$.} Scalar condensation may happen at a sufficiently low temperature. Numerical solutions of hyperbolic black holes with scalar hair were obtained \cite{Dias:2010ma,Belin:2013dva}, and a class of analytic solutions of hyperbolic black holes with scalar hair was found in~\cite{Ren:2019lgw}. 

The first type of instability has not been analytically calculated so far. We need to solve the Klein-Gordon equation at $\omega=0$. A sufficiently large charge of the scalar field will lead to the first type of instability, as well as the second one. Therefore, we need to examine which comes first. We study two gravitational systems dual to the $\mathcal{N}=4$ super-Yang-Mills theory at finite density.

The first system we study is the RN-AdS$_5$ black hole. To examine the phase transition of R\'{e}nyi entropies, we study the instability of the extremal hyperbolic RN-AdS$_5$ black hole. The quantum critical point is at
\begin{equation}
\tilde{q}-\nu-\frac{\Delta_\pm\pm 1}{2}=n_\pm\,,
\label{eq:zero-RN}
\end{equation}
where $\tilde{q}$ is proportional to the charge of the scalar field, $\nu$ is the IR scaling exponent, $\Delta_\pm$ is the scaling dimension of the dual scalar operator, and $n_\pm$ is a non-negative integer; the subscript ``$+$'' is for the standard quantization, and the subscript ``$-$'' is for the alternative quantization. The first type of instability happens when a zero mode determined by~\eqref{eq:zero-RN} is satisfied. The second type of instability happens when the IR scaling exponent $\nu$ becomes imaginary.

On one hand, the condition for a zero mode requires a sufficiently large $\tilde{q}$. On the other hand, the IR scaling exponent $\nu$ becomes imaginary at sufficiently large $\tilde{q}$. As we increase $\tilde{q}$, we find that the IR instability always comes before the zero mode condition~\eqref{eq:zero-RN} is met for the hyperbolic RN-AdS$_5$ black hole.

The second system we study is the Gubser-Rocha model~\cite{Gubser:2009qt}, in which the black hole is dilatonic and has zero entropy at zero temperature, in contrast to the RN-AdS black hole. The IR geometry of the extremal hyperbolic charged black hole is conformal to AdS$_2\times\mathbb{H}_3$. The IR scaling exponent is always real, and thus there is no IR instability. We study a massless charged scalar field in the bulk to perturb the hyperbolic black hole, and look for a zero mode. The quantum critical point is at
\begin{equation}
\tilde{q}-\nu-3=2n\,,
\label{eq:zero-GR}
\end{equation}
where $\tilde{q}$ is proportional to the charge of the scalar field, $\nu$ is the IR scaling exponent, and $n$ is a non-negative integer. The IR scaling dimension $\nu$ never becomes imaginary in this system, and thus only a zero mode causes an instability.

This paper is organized as follows. In section~\ref{sec:RN}, we study the instabilities of the hyperbolic RN-AdS$_5$ black hole, and analytically solve the zero mode and the Green's function near the quantum critical point. In section~\ref{sec:GR}, we study the instabilities of the hyperbolic charged black hole in the Gubser-Rocha model, and analytically solve the zero mode. Finally, we summarize and discuss some open questions. In appendix~\ref{sec:FS}, we present an analytic solution of the zero mode for the Dirac equation in a closely related system. In appendix~\ref{sec:math}, we give some mathematical notes.

\subsection{From R\'{e}nyi entropies to hyperbolic black holes}
We review the relation between R\'{e}nyi entropies and hyperbolic black holes very briefly, but it is sufficient for the purpose of this paper. For details, see~\cite{Casini:2011kv,Hung:2011nu}. Phase transitions of R\'{e}nyi entropies with the entangling surface being a sphere are studied in terms of hyperbolic black holes. For hyperbolic black holes and their holographic properties, see~\cite{Emparan:1998he,Birmingham:1998nr,Emparan:1999gf} for example.

Consider a quantum field theory in a state described by a density matrix $\rho$, and divide the system into two parts, A and B. The reduced density matrix for the subsystem A is $\rho_A=\text{Tr}_B\rho$. The R\'{e}nyi entropy is defined by
\begin{equation}
S_n=\frac{1}{1-n}\log\textrm{Tr}[\rho_A^n]\,.
\end{equation}
The entanglement entropy $S_{EE}$ can be obtained by taking the $n\to 1$ limit of the R\'{e}nyi entropy: $S_{EE}=\lim_{n\to 1}S_n=-\text{Tr}\rho_A\log\rho_A$. The charged R\'{e}nyi entropy is defined by \cite{Belin:2013uta}
\begin{equation}
S_n=\frac{1}{1-n}\log\textrm{Tr}\left[\rho_A\frac{\mu Q_A}{\mathcal{N}_A(\mu)}\right]^n,
\end{equation}
where $\mu$ is the entanglement chemical potential, and $Q_A$ measures the amount of charge in the subsystem A, and $\mathcal{N}_A(\mu)\equiv\text{Tr}[\rho_Ae^{\mu Q_A}]$ is a normalization factor.

Suppose we want to calculate the R\'{e}nyi entropies of a CFT with a gravity dual, and the entangling surface is a sphere of radius $R$. By a conformal mapping, the R\'{e}nyi entropy is related to the free energy of a hyperbolic black hole:
\begin{equation}
S_n(\mu)=\frac{n}{1-n}\frac{1}{T_0}(F(T_0)-F(T_0/n))=\frac{n}{1-n}\frac{1}{T_0}\int_{T_0/n}^{T_0}S_\text{therm}(T,\mu)dT\,,
\end{equation}
where $T_0=\frac{1}{2\pi R}$ is the temperature of a zero-mass hyperbolic black hole, $S_\text{therm}$ is the thermal entropy of the hyperbolic black hole, and $S_\text{therm}=-\partial F/\partial T$.

\section{Zero modes from the hyperbolic RN-AdS$_5$ black hole\label{sec:RN}}
\subsection{Hyperbolic RN-AdS$_5$ black hole}
We consider the action
\begin{equation}
S=\int d^5x\sqrt{-g}\left[\frac{1}{2\kappa^2}\left(R+\frac{12}{L^2}\right)-\frac{1}{4g^2}F_{\mu\nu}F^{\mu\nu}-\frac{1}{2}|(\nabla_\mu-iqA_\mu)\Phi|^2-V(\Phi)\right],
\end{equation}
where the potential of the charged scalar field is a mass term $V(|\Phi|)=\frac{1}{2}m^2|\Phi|^2$.  We first solve the background solution with $\Phi=0$, and then perturb the system by $\Phi$. The metric for the hyperbolic RN-AdS$_5$ black hole in Poincar\'{e} coordinates is
\begin{equation}
ds^2=\frac{L^2}{z^2}\left(-f(z)dt^2+K^{-2}\tsp d\mathbb{H}_3^2+\frac{dz^2}{f(z)}\right),\label{eq:metric-RN}
\end{equation}
where $d\mathbb{H}_3^2$ is the metric for the 3-dimensional hyperbolic space with unit radius.
The boundary metric is
\begin{equation}
ds_\partial^2=-dt^2+ds_3^2\,,
\end{equation}
where the spatial part is the metric for the hyperbolic space with radius $K^{-1}$:
\begin{equation}
ds_3^2 =\frac{dr^2}{1+K^2r^2}+r^2d\Omega_2^2=K^{-2}(d\rho^2+\sinh^2\rho d\Omega_2^2)=K^{-2}\tsp d\mathbb{H}_3^2\,.
\end{equation}
The relation between the $r$ and $\rho$ coordinates is $Kr=\sinh\rho$.

The solution to the metric is~\eqref{eq:metric-RN} with
\begin{equation}
f=1-K^2z^2-\left(1-K^2z_h^2+\frac{z_h^2\mu^2}{\gamma^2}\right)\left(\frac{z}{z_h}\right)^4+\frac{z_h^2\mu^2}{\gamma^2}\left(\frac{z}{z_h}\right)^6,
\label{eq:metric-RNf}
\end{equation}
where $z_h$ is the horizon radius, $\mu$ is the chemical potential, and $\gamma$ is a dimensionless quantity
\begin{equation}
\gamma^2=\frac{3g^2L^2}{2\kappa^2}.
\end{equation}
The solution to the gauge field $A=A_tdt$ is
\begin{equation}
A_t=\mu\left(1-\frac{z^2}{z_h^2}\right).
\end{equation}
We can set $L=1$, $2\kappa^2=1$, and $g=1$, and thus $\gamma^2=3$. The metric with~\eqref{eq:metric-RNf} has three dimensionful parameters: the horizon radius $z_h$, the chemical potential $\mu$, and the curvature of the hyperbolic space $K$. Since a scale invariant theory only depends on dimensionless parameters, we set $z_h=1$, and then $\mu$ and $K$ are two independent parameters.
The temperature is
\begin{equation}
T=\frac{f'(z_h)}{4\pi}=\frac{6-3K^2-\mu^2}{6\pi}.
\end{equation}
After we take the zero temperature limit, only one parameter will remain, and we choose it to be $K$. Consequently, the extremal black hole is obtained by setting
\begin{equation}
\mu=\sqrt{3(2-K^2)}\,.
\end{equation}
We can see that the physical range of $K^2$ is $0<K^2<2$ for the hyperbolic RN-AdS$_5$ black hole.

\subsection{Analytic solution of the Klein-Gordon equation at $\omega=0$}
We solve the Klein-Gordon equation
\begin{equation}
[(\nabla^\mu-iqA^\mu)(\nabla_\mu-iqA_\mu)-m^2]\Phi=0
\end{equation}
for the charged scalar field $\Phi$ in the above background to obtain the Green's function of the dual scalar operator in the CFT. After the separation of variables
\begin{equation}
\Phi(z,x^\mu)\sim e^{-i\omega t}Y(\sigma)\phi(z)\,,
\end{equation}
where $Y(\sigma)$ satisfies $\hat{\nabla}^2Y=-\lambda Y$ with $\hat{\nabla}^2$ being the Laplacian on $\mathbb{H}_3$. For a normalizable mode on $\mathbb{H}_3$, we have $\lambda\geq 1$. The equation of motion for $\phi$ is
\begin{equation}
\phi''+\left(\frac{f'}{f}-\frac{3}{z}\right)\phi'+\left(\frac{(\omega+qA_t)^2}{f^2}-\frac{\lambda K^2}{f}
-\frac{m^2}{z^2f}\right)\phi=0\,,\label{eq:KG-RN}
\end{equation}
where we assume $q>0$, and $m^2$ is above the BF bound: $m^2\geq m_\text{BF}^2=-4$. If we take the $K\to 0$ limit and replace $\lambda K^2$ with $\mathbf{k}^2$, we obtain the Klein-Gordon equation for the planar RN-AdS$_5$ black hole with $\mathbf{k}$ being the spatial momentum. Replacing $K^2$ with $-K^2$ gives the equation for the spherical RN-AdS$_5$ black hole.

With the infalling boundary condition at the horizon, the asymptotic behavior of the scalar field near the AdS boundary is\footnote{When $\Delta_+-\Delta_-=2n$, where $n=1$, $2$, $\cdots$, there are extra terms $bz^{\Delta_+}\ln z\,(1+\cdots)$. When $\Delta_+=\Delta_-$, $\phi=Az^2\ln z+Bz^2+\cdots$.}
\begin{equation}
\phi=Az^{\Delta_-}(1+\cdots)+Bz^{\Delta_+}(1+\cdots)\,.
\end{equation}
The retarded Green's function is
\begin{equation}
G=\frac{B}{A}\,.
\end{equation}
We have ignored an unimportant prefactor. When $-4\leq m^2\leq -3$, there is an alternative quantization, by which the Green's function is $G=A/B$ \cite{Klebanov:1999tb}.

To study the instability near a quantum critical point, we need to solve the Green's function near $\omega=0$. When $\omega=0$, we can solve $\phi$ in terms of hypergeometric equations. The general solution of $\phi$ for~\eqref{eq:KG-RN} at $\omega=0$ is\footnote{When $\Delta_+$ is an integer, the two hypergeometric functions in~\eqref{eq:sol-RN} are linearly dependent. We can choose another two linearly independent solutions as~\eqref{eq:sol2-RN} in appendix~\ref{sec:math}.}
\begin{align}
\phi(z)= &C_1z^{\Delta_-}\frac{(1-z^2)^{-1/2+\nu}}{((2-K^2)z^2+1)^{-1/2+\nu+\Delta_-/2}}\times\nonumber\\
&\qquad\times{_2F_1}\left(\frac{\Delta_--1}{2}+\nu-\tilde{q},\,\frac{\Delta_--1}{2}+\nu+\tilde{q};\,
\Delta_--1;\,\frac{(3-K^2)z^2}{(2-K^2)z^2+1}\right)\nonumber\\
+ &C_2z^{\Delta_+}\frac{(1-z^2)^{-1/2+\nu}}{((2-K^2)z^2+1)^{-1/2+\nu+\Delta_+/2}}\times\nonumber\\
&\qquad\times{_2F_1}\left(\frac{\Delta_+-1}{2}+\nu-\tilde{q},\,\frac{\Delta_+-1}{2}+\nu+\tilde{q};\,
\Delta_+-1;\,\frac{(3-K^2)z^2}{(2-K^2)z^2+1}\right),\label{eq:sol-RN}
\end{align}
where
\begin{align}
\tilde{q} &=\frac{2-K^2}{3-K^2}\frac{\sqrt{3}}{2}q\,,\label{eq:qtil-RN}\\
\nu &=\frac{1}{2(3-K^2)}\sqrt{(3-K^2)(m^2+\lambda K^2+3-K^2)-3(2-K^2)q^2}\,,\label{eq:nu-RN}\\
\Delta_\pm &=2\pm\sqrt{m^2+4}\,.\label{eq:Delta-RN}
\end{align}
This is the key result that enables us to extract the analytic solution for zero modes.

\subsection{Matching the IR and UV data}
We expect that the Green's function near the critical point can be obtained by the perturbation of small $\omega$ around the exact solution. However, for the extremal geometry, the horizon $z=1$ is an irregular singularity in~\eqref{eq:KG-RN}. When it is sufficiently close to the extremal horizon, $\omega$-dependent terms cannot be treated as small perturbations no matter how small $\omega$ is. In~\cite{Faulkner:2009wj}, a systematic method is developed for treating extremal black hole systems. Applications of this method for studying quantum critical systems from planar charged black holes include~\cite{Iqbal:2011aj,Ren:2012hg,Alishahiha:2012ad}, where analytic solutions for zero modes are obtained in~\cite{Ren:2012hg,Alishahiha:2012ad}. This section generalizes~\cite{Ren:2012hg}, and the next section generalizes~\cite{Alishahiha:2012ad} to corresponding hyperbolic cases.

\begin{figure}
  \centering
  \includegraphics[]{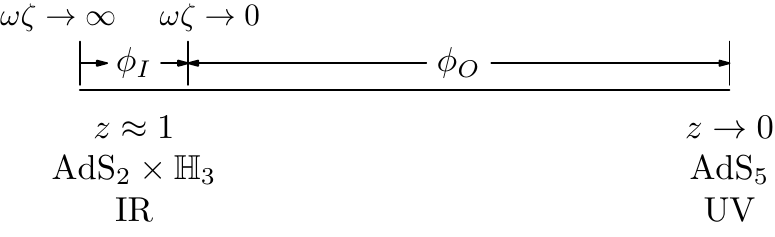}
  \caption{\label{fig:match} The inner (near horizon) and outer regions, where the solutions of the Klein-Gordon equation are denoted by $\phi_I$ and $\phi_O$, respectively.}
\end{figure}

We divide the geometry into inner and outer regions, as shown in figure~\ref{fig:match}. The inner region refers to the IR (near horizon) geometry, in which the Klein-Gordon equation with arbitrary $\omega$ can be exactly solved as~\eqref{eq:phiin-RN} below. The outer region refers to the remaining geometry, in which we can make perturbations for small $\omega$. Then we need to match the inner and outer regions.

The IR geometry is obtained by taking the $z\to 1$ limit, in which we have
\begin{equation}
f=4(3-K^2)(1-z)^2,\qquad A_t=2\sqrt{3(2-K^2)}(1-z)\,.
\end{equation}
After the change of variables
\begin{equation}
\zeta=\frac{1}{4(3-K^2)(1-z)}\,,\qquad L_2=\frac{1}{\sqrt{4(3-K^2)}}\,,
\end{equation}
the metric becomes
\begin{equation}
ds^2=\frac{L_2^2}{\zeta^2}(-dt^2+d\zeta^2)+K^{-2}d\mathbb{H}_3^2\,.
\end{equation}
Therefore, the IR geometry is AdS$_2\times\mathbb{H}_3$ with $L_2$ being the AdS$_2$ radius. The solution to the Klein-Gordon equation in this background with the infalling boundary condition at the horizon is\footnote{The infalling wave in terms of the coordinate $\zeta$ is $e^{i\omega\zeta}$ as $\zeta\to\infty$.}
\begin{equation}
\phi\sim W_{iq',\nu}(-2i\omega\zeta)\,,\label{eq:phiin-RN}
\end{equation}
where
\begin{equation}
q'=\frac{\sqrt{3(2-K^2)}}{2(3-K^2)}q\,,
\end{equation}
and $W_{\lambda,\mu}(x)$ is a Whittaker function with the following asymptotic behavior:
\begin{equation}
W_{\lambda,\mu}(x)\sim e^{-x/2}x^\lambda(1+\cdots)\,,\qquad |x|\to\infty\,.
\end{equation}
By expanding~\eqref{eq:phiin-RN} at $\omega\zeta\to 0$, we obtain
\begin{equation}
\phi=\zeta^{1/2-\nu}(1+\cdots)+\mathcal{G}(\omega)\zeta^{1/2+\nu}(1+\cdots)\,.
\end{equation}
The IR Green's function at zero temperature is
\begin{equation}
\mathcal{G}(\omega)=\frac{\Gamma(-2\nu)\Gamma\bigl(\frac{1}{2}+\nu-iq'\bigr)}
{\Gamma(2\nu)\Gamma\bigl(\frac{1}{2}-\nu-iq'\bigr)}(-2i\omega)^{2\nu}.\label{eq:gir}
\end{equation}

In the outer region, the solution at small $\omega$ can be written as
\begin{equation}
\phi(z)=\eta_+(z)+\mathcal{G}_k(\omega)\eta_-(z)\,,
\end{equation}
where
\begin{equation}
\eta_\pm=\eta_\pm^{(0)}+\omega\eta_\pm^{(1)}+\mathcal{O}(\omega^2)\,,
\end{equation}
where $\eta_+^{(0)}$ is normalizable and $\eta_-^{(0)}$ is nonnormalizable at the horizon. At the leading order, the asymptotic behavior near the horizon $z=1$ is
\begin{equation}
\eta_\pm^{(0)}\to\zeta^{1/2\mp\nu}=[4(3-K^2)(1-z)]^{-1/2\pm\nu},\label{eq:etah}
\end{equation}
which we also use to fix the normalization of the solution. The asymptotic behavior near the AdS boundary $z=0$ is
\begin{equation}
\eta_\pm^{(0)}\to a_\pm^{(0)}z^{\Delta_-}+b_\pm^{(0)}z^{\Delta_+}.\label{eq:etab}
\end{equation}
The Green's function to the first order in $\omega$ is \cite{Faulkner:2009wj}
\begin{equation}
G(\omega,k)=\frac{b_+^{(0)}+\omega b_+^{(1)}+O(\omega^2)+{\cal G}(\omega)\bigl(b_-^{(0)}+\omega b_-^{(1)}+O(\omega^2)\bigr)}
{a_+^{(0)}+\omega a_+^{(1)}+O(\omega^2)+{\cal G}(\omega)\bigl(a_-^{(0)}+\omega a_-^{(1)}+O(\omega^2)\bigr)}\,.\label{eq:green}
\end{equation}
The analytic solution of $\phi$ at $\omega=0$ gives the leading order of the Green's function. By perturbation around $\omega=0$, we can obtain the higher-order coefficients. (For a neutral scalar, the first order terms in $\omega$ are zero, and then we need to expand to the second order.) The Green's function can be generalized to nonzero temperature when $T<<\mu$ (chemical potential) by replacing the IR Green's function $\mathcal{G}(\omega)$ with the finite temperature solution in AdS$_2$.

With the solution~\eqref{eq:sol-RN} at hand, we can obtain the analytic solution of $a_\pm^{(0)}$ and $b_\pm^{(0)}$. The asymptotic behavior of $\phi$ near the horizon $z\to 1$ is
\begin{align}
\phi\to \,&[2(1-z)]^{-1/2-\nu}\times\nonumber\\
&\times\left(\frac{C_1\,3^{1/2+\nu-\Delta_-/2}\Gamma(\Delta_--1)\Gamma(2\nu)}
{\Gamma\bigl(\frac{\Delta_--1}{2}+\nu+\tilde{q}\bigr)
\Gamma\bigl(\frac{\Delta_--1}{2}+\nu-\tilde{q}\bigr)}
+\frac{C_2\,3^{1/2+\nu-\Delta_+/2}\Gamma(\Delta_+-1)\Gamma(2\nu)}
{\Gamma\bigl(\frac{\Delta_+-1}{2}+\nu+\tilde{q}\bigr)
\Gamma\bigl(\frac{\Delta_+-1}{2}+\nu-\tilde{q}\bigr)}\right)\nonumber\\
+ \,&[2(1-z)]^{-1/2+\nu}\times\nonumber\\
&\times\left(\frac{C_1\,3^{1/2-\nu-\Delta_-/2}\Gamma(\Delta_--1)\Gamma(-2\nu)}
{\Gamma\bigl(\frac{\Delta_--1}{2}-\nu+\tilde{q}\bigr)
\Gamma\bigl(\frac{\Delta_--1}{2}-\nu-\tilde{q}\bigr)}
+\frac{C_2\,3^{1/2-\nu-\Delta_+/2}\Gamma(\Delta_+-1)\Gamma(-2\nu)}
{\Gamma\bigl(\frac{\Delta_+-1}{2}-\nu+\tilde{q}\bigr)
\Gamma\bigl(\frac{\Delta_+-1}{2}-\nu-\tilde{q}\bigr)}\right).
\end{align}
The asymptotic behavior of $\phi$ near the boundary $z\to 0$ is
\begin{equation}
\phi\to C_1z^{\Delta_-}+C_2z^{\Delta_+}.
\end{equation}
By~\eqref{eq:etah} and~\eqref{eq:etab}, the solutions of $a_\pm^{(0)}$ and $b_\pm^{(0)}$ are
\begin{align}
a_+^{(0)} &=\frac{\nu}{\sqrt{m^2+4}}\cdot\dfrac{2^{1/2+\nu}(3-K^2)^{1+2\nu-\Delta_+/2}\Gamma(\Delta_+-1)\Gamma(2\nu)}
{\Gamma\bigl(\frac{\Delta_+-1}{2}+\nu+\tilde{q}\bigr)
\Gamma\bigl(\frac{\Delta_+-1}{2}+\nu-\tilde{q}\bigr)},\label{eq:a0p-RN}\\[3pt]
b_+^{(0)} &=-\frac{\nu}{\sqrt{m^2+4}}\cdot\dfrac{2^{1/2+\nu}(3-K^2)^{1+2\nu-\Delta_-/2}\Gamma(\Delta_--1)\Gamma(2\nu)}
{\Gamma\bigl(\frac{\Delta_--1}{2}+\nu+\tilde{q}\bigr)
\Gamma\bigl(\frac{\Delta_--1}{2}+\nu-\tilde{q}\bigr)},\label{eq:b0p-RN}\\[3pt]
a_-^{(0)} &=-\frac{\nu}{\sqrt{m^2+4}}\cdot\dfrac{2^{1/2-\nu}(3-K^2)^{1-2\nu-\Delta_+/2}\Gamma(\Delta_+-1)\Gamma(-2\nu)}
{\Gamma\bigl(\frac{\Delta_+-1}{2}-\nu+\tilde{q}\bigr)
\Gamma\bigl(\frac{\Delta_+-1}{2}-\nu-\tilde{q}\bigr)},\\[3pt]
b_-^{(0)} &=\frac{\nu}{\sqrt{m^2+4}}\cdot\dfrac{2^{1/2-\nu}(3-K^2)^{1-2\nu-\Delta_-/2}\Gamma(\Delta_--1)\Gamma(-2\nu)}
{\Gamma\bigl(\frac{\Delta_--1}{2}-\nu+\tilde{q}\bigr)
\Gamma\bigl(\frac{\Delta_--1}{2}-\nu-\tilde{q}\bigr)}.
\end{align}
It can be checked that
\begin{equation}
a_+^{(0)}b_-^{(0)}-a_-^{(0)}b_+^{(0)}=\frac{\nu}{\sqrt{m^2+4}}
\end{equation}
 is satisfied.

\subsection{Two types of instabilities}
A zero mode is defined by a pole of the Green's function at $\omega=0$. Therefore, zero modes are determined by $a_+^{(0)}=0$ for the standard quantization, and $b_+^{(0)}=0$ for the alternative quantization with $-4\leq m^2\leq -3$. By~\eqref{eq:a0p-RN} and~\eqref{eq:b0p-RN}, we obtain an analytic solution for the zero modes at the quantum critical point:
\begin{equation}
\tilde{q}-\nu-\frac{\Delta_\pm-1}{2}=n_\pm\,,\qquad n_\pm=0,1,2,\cdots,
\label{eq:zero-mode-RN}
\end{equation}
where $\tilde{q}$, $\nu$, and $\Delta_\pm$ are given by~\eqref{eq:qtil-RN}, \eqref{eq:nu-RN}, and \eqref{eq:Delta-RN}, respectively; the subscript ``$+$'' is for the standard quantization, and the subscript ``$-$'' is for the alternative quantization. The most unstable zero mode is at $n_\pm=0$ and $\lambda=1$.

\begin{figure}
  \centering
  \includegraphics[]{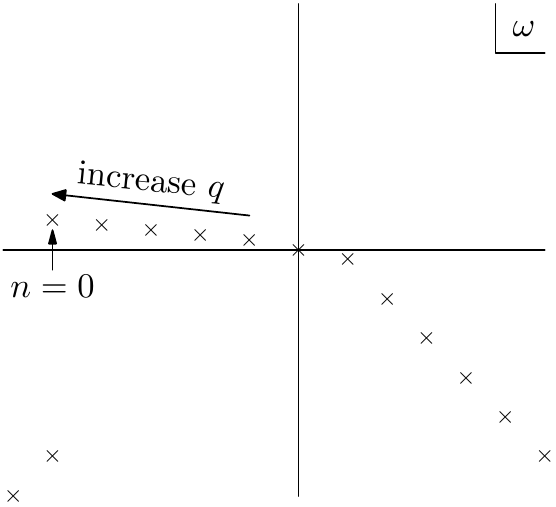}
  \caption{\label{fig:qnm} Schematic plot of the poles of the Green's function when $q$ is large.}
\end{figure}

It is helpful to understand the zero modes by looking at the poles of the Green's function at arbitrary $\omega$. Numerical calculations suggest the following features, as illustrated in figure~\ref{fig:qnm}. At $q=0$ and $\lambda=1$, all the poles of the Green's function are in the lower half $\omega$-plane, and not close to $\omega=0$. As we increase $q$, there are more and more poles moving across the origin to the upper half $\omega$-plane, and the first one is labeled by $n=0$. If we start from a large $q$ and increase $\lambda$, the poles on the upper half $\omega$-plane will move across the origin to the lower half $\omega$-plane. Moreover, $\lambda=1$ corresponds to the most unstable mode, i.e., if there are poles in the upper half $\omega$-plane when $\lambda>1$, there are no less poles in the upper half $\omega$-plane when $\lambda=1$. Therefore, the onset of the instability signaled by a zero mode happens when the first pole moves across the origin with $\lambda=1$ as we increase $q$, provided that the IR instability does not exist.

\begin{figure}
\begin{minipage}[t]{\textwidth}
\centering
  \includegraphics[width=0.32\textwidth]{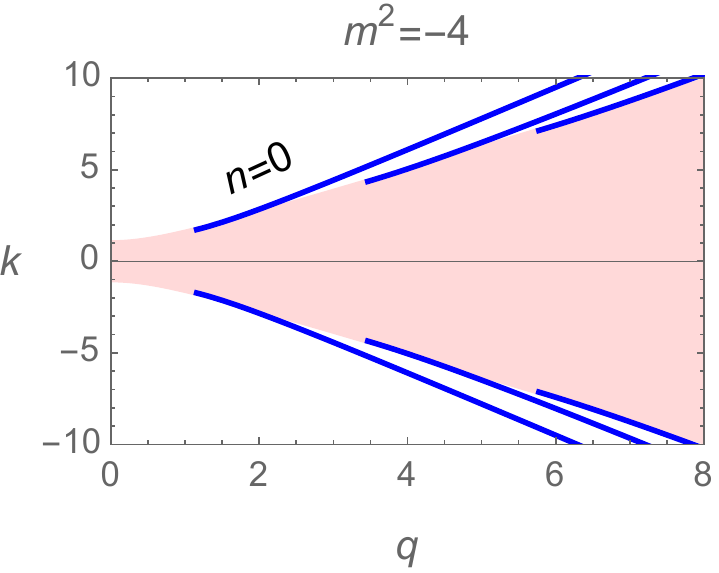}\;
  \includegraphics[width=0.32\textwidth]{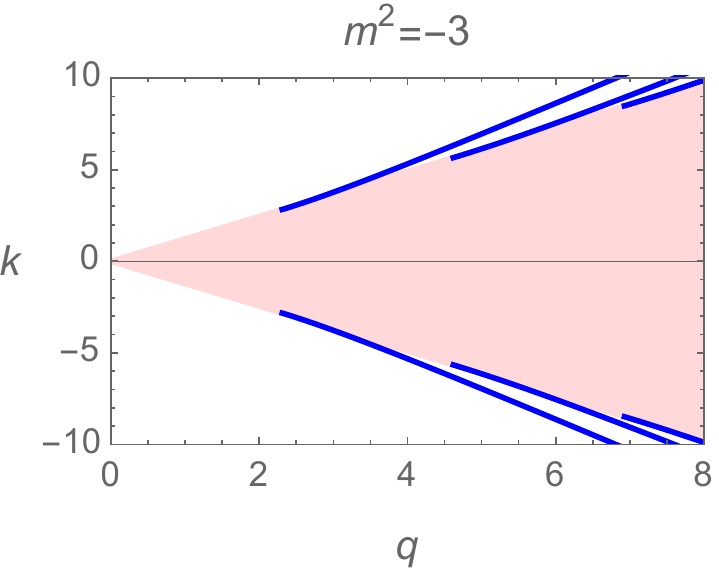}\;
  \includegraphics[width=0.32\textwidth]{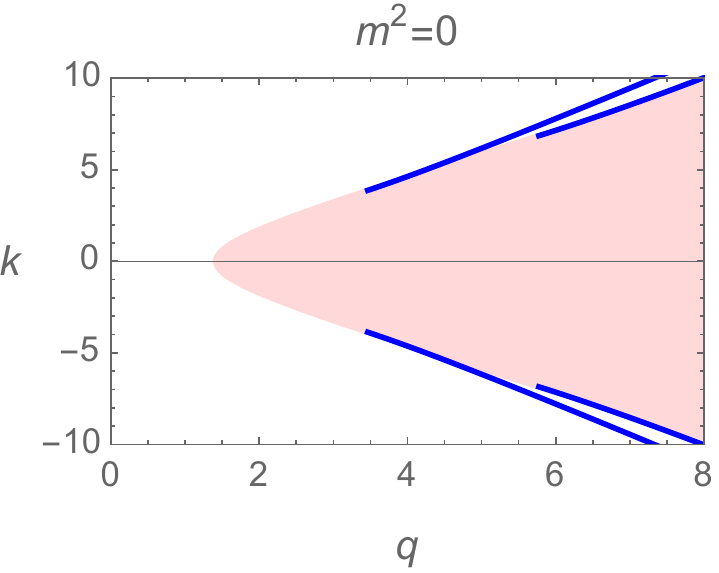}\;
  \caption{\label{fig:kqs} Phase diagram for the standard quantization. We take $K=1$, and the vertical axis is $k\equiv \sqrt{\lambda-1}$. The solid lines correspond to zero modes. The region with IR instability is shaded, and will move to the right as we increase $m^2$. In the right plot, the tip of the shaded region corresponds to a bifurcating critical point at $k=0$.}
\end{minipage}\\[15pt]
\begin{minipage}[t]{\textwidth}
\centering
  \includegraphics[width=0.32\textwidth]{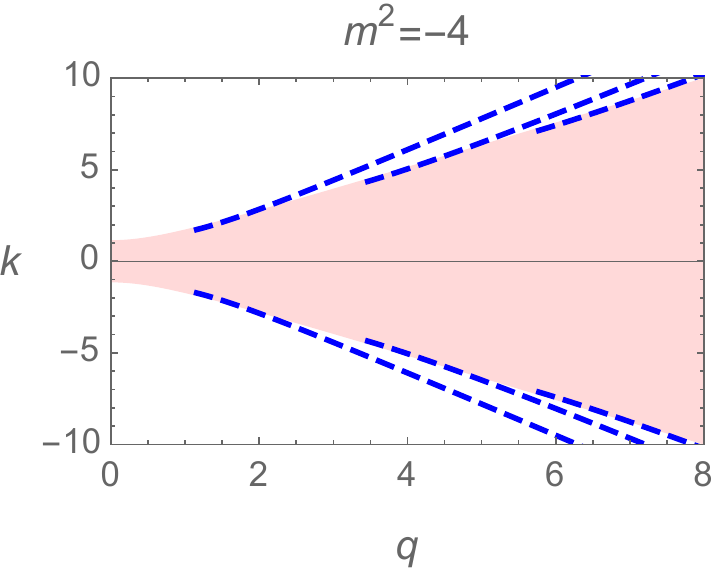}\;
  \includegraphics[width=0.32\textwidth]{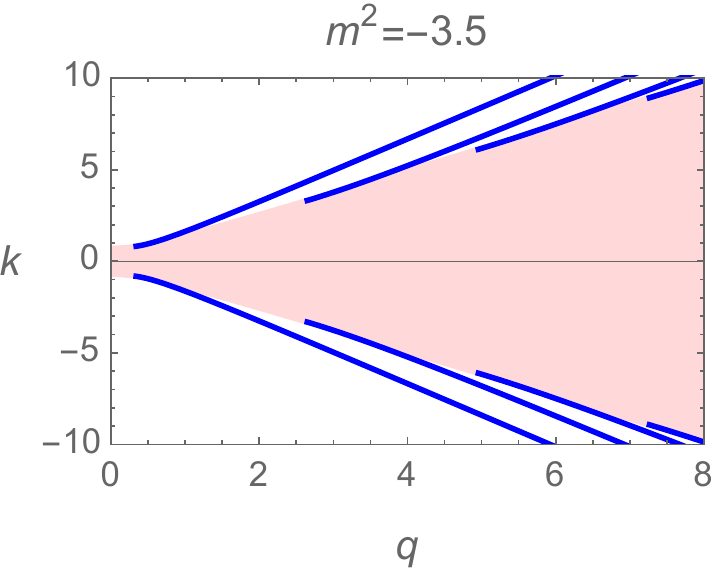}\;
  \includegraphics[width=0.32\textwidth]{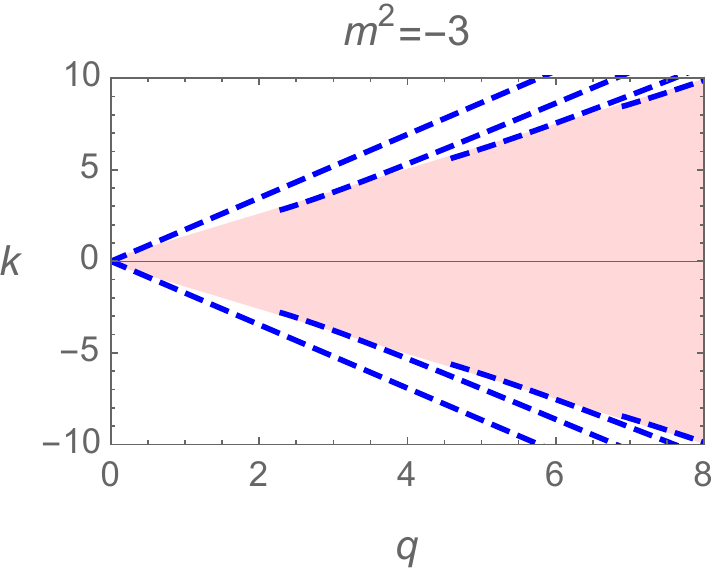}
  \caption{\label{fig:kqa} Phase diagram for the alternative quantization. We take $K=1$, and the vertical axis is $k\equiv \sqrt{\lambda-1}$. The region with IR instability is shaded.}
\end{minipage}
\end{figure}

To have at least one zero mode, $\tilde{q}$ must be sufficiently large. If $\tilde{q}$ is large, the IR scaling exponent $\nu$ may become imaginary, causing an IR instability. Therefore, we need to examine which type of the instabilities comes first. If the condition~\eqref{eq:zero-mode-RN} is satisfied with a real $\nu$, it is necessary that
\begin{equation}
\tilde{q}\geq \frac{\Delta_\pm-1}{2}=\frac{1\pm\sqrt{m^2+4}}{2}\equiv \tilde{q}_{c\pm}\,.
\label{eq:qc-RN}
\end{equation}
We can check that $\nu(\tilde{q}_{c\pm})$ is always imaginary except for $q=0$, which does not give a normal mode. Consequently, the IR instability always comes first. In figures~\ref{fig:kqs} and~\ref{fig:kqa}, zero modes are plotted as solid lines for various parameters, and the IR instability happens in the shaded region. We can see that the $\lambda=1$ line never intersects with the solid lines.

The condition for the zero modes~\eqref{eq:zero-mode-RN} is non-universal, i.e., it depends on the specific model. Therefore, we expect that this type of instability is important when the IR instability is absent.

\begin{figure}
  \centering
  \includegraphics[]{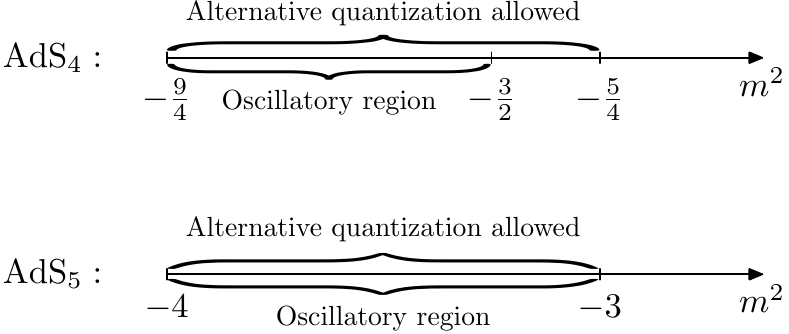}
  \caption{\label{fig:m2osc} A slight difference between AdS$_4$ and AdS$_5$. We set the AdS radius $L=1$. In AdS$_4$, there is an interval for $m^2$ in which the alternative quantization is allowed without IR instability. In AdS$_5$, such interval does not exist.}
\end{figure}

\subsection{Adding a double-trace deformation}
Recall that the Green's function for the standard quantization is $G=B/A$. The zero modes can be achieved by turning on another parameter $\kappa_+$, which describes a double trace deformation in the boundary CFT:
\begin{equation}
\frac{\kappa_+}{2}\int d^5x\,\mathcal{O}^2,
\end{equation}
where $\langle\mathcal{O}\rangle=B$ \cite{Iqbal:2011aj, Witten:2001ua}. The Green's function becomes
\begin{equation}
G^{(\kappa_+)}=\frac{1}{G^{-1}+\kappa_+}\,.
\end{equation}
The Green's function at the leading order in $\omega$ is
\begin{equation}
G(\omega,k)=\frac{b_+^{(0)}+{\cal G}_k(\omega)b_-^{(0)}}
{a_+^{(0)}+\kappa_+b_+^{(0)}+{\cal G}_k(\omega)(a_-^{(0)}+\kappa_+b_-^{(0)})}\,.\label{eq:Ghybri}
\end{equation}
The boundary condition for a pole in the Green's function at $\omega=0$ is $a_+^{(0)}+\kappa_+b_+^{(0)}=0$. Thus, by~\eqref{eq:a0p-RN} and~\eqref{eq:b0p-RN}, the critical value of $\kappa_+$ is
\begin{equation}
\kappa_c=3^{-\sqrt{m^2+4}}\cdot
\frac{\Gamma\bigl(\frac{\Delta_--1}{2}+\nu+\tilde{q}\bigr)
\Gamma\bigl(\frac{\Delta_--1}{2}+\nu-\tilde{q}\bigr)\Gamma(\Delta_+-1)}
{\Gamma\bigl(\frac{\Delta_+-1}{2}+\nu+\tilde{q}\bigr)
\Gamma\bigl(\frac{\Delta_+-1}{2}+\nu-\tilde{q}\bigr)\Gamma(\Delta_--1)}\,.\label{eq:kappa}
\end{equation}
Besides this type of instability, another type of instability is when the IR scaling exponent $\nu$ becomes imaginary. There will be an intricate phase diagram if we take into account both types of instability.

Qualitative features are expected to be similar in AdS$_4$. In the AdS$_4$ case, the parameter range of $m^2$ when the alternative quantization is allowed is slightly larger than the interval when there is IR instability. In AdS$_5$, the two intervals coincide. See figure~\ref{fig:m2osc}.

\section{Zero modes from the hyperbolic Gubser-Rocha model\label{sec:GR}}
\subsection{Hyperbolic 2-charge black hole in AdS$_5$}

The second system we study is the Gubser-Rocha model~\cite{Gubser:2009qt}, in which the black hole is dilatonic and has zero entropy at zero temperature, in contrast to the RN-AdS black hole. It is called the 2-charge black hole in AdS$_5$ in the following sense. In five-dimensional maximal gauged supergravity, two charges of the three $U(1)$ subgroups of the $SO(6)$ gauge group are nonzero and equal, and the third is zero. We will show that the IR geometry of the hyperbolic 2-charge black hole at zero temperature is conformal to AdS$_2\times\mathbb{H}_3$. The IR scaling exponent is always real, and thus there is no IR instability.

The 2-charge black hole in AdS$_5$ is determined by
\begin{equation}
S=\int d^5x\sqrt{-g}\left[R-\frac{1}{4}e^{\frac{2}{\sqrt{6}}\alpha}F_{\mu\nu}^2-\frac{1}{2}(\partial_\mu\alpha)^2+\frac{1}{L^2}(8e^{\frac{1}{\sqrt{6}}\alpha}+4e^{-\frac{2}{\sqrt{6}}\alpha})\right],
\end{equation}
which is from a consistent truncation of the type IIB supergravity with three U(1) charges $Q_1=Q_2=Q$ and $Q_3=0$. The metric for the black hole solution is
\begin{gather}
ds^2=e^{2\mathcal{A}}(-hdt^2+K^{-2}d\mathbb{H}_3^2)+\frac{e^{2\mathcal{B}}}{h}dr^2,\\
\mathcal{A}=\ln\frac{r}{L}+\frac{1}{3}\ln\left(1+\frac{Q^2}{r^2}\right),\qquad \mathcal{B}=-\ln\frac{r}{L}-\frac{2}{3}\ln\left(1+\frac{Q^2}{r^2}\right),\nonumber\\
h=1-\frac{K^2L^4}{r^2+Q^2}-\frac{(r_h^2+Q^2)(r_h^2+Q^2-K^2L^4)}{(r^2+Q^2)^2},\nonumber
\end{gather}
and the gauge field $A=A_tdt$ and the dilaton are
\begin{gather}
A_t=\sqrt{\frac{2(r_h^2+Q^2-K^2L^4)}{r_h^2+Q^2}}\frac{Q}{L}\left(1-\frac{r_h^2+Q^2}{r^2+Q^2}\right),\\
\alpha=\frac{2}{\sqrt{6}}\ln\left(1+\frac{Q^2}{r^2}\right),
\end{gather}
where $r_h$ is the horizon radius, $Q$ is a parameter related to the chemical potential, $K$ is the curvature of the hyperbolic space, and $L=1$ is the AdS radius.

The temperature of this black hole is
\begin{equation}
T=\left.\frac{|h'|e^{\mathcal{A}-\mathcal{B}}}{4\pi}\right|_{r=r_h}=\frac{r_h}{2\pi}\left(2-\frac{K^2}{r_h^2+Q^2}\right).
\end{equation}
We will consider the zero temperature black hole solution, which is at $r_h=0$. The ``horizon'' (IR limit) for this black hole is at $r=0$, which is a spacetime singularity. However, the spacetime singularity is cloaked by a horizon at finite temperature. It has a ten-dimensional lift; see~\cite{Gubser:2009qt,Cvetic:1999xp}.

To obtain the Green's function for a scalar operator in the dual CFT, we solve the Klein-Gordon equation for a charged scalar field $\Phi$. After the separation of variables
\begin{equation}
\Phi(r,x^\mu)\sim e^{-i\omega t}Y(\sigma)\phi(r)\,,
\end{equation}
where $Y(\sigma)$ satisfies $\hat{\nabla}^2Y=-\lambda Y$ with $\hat{\nabla}^2$ being the Laplacian on $\mathbb{H}_3$. For a normalizable mode on $\mathbb{H}_3$, we have $\lambda\geq 1$. The equation of motion for $\phi$ is
\begin{equation}
\phi''+\left(4\mathcal{A}'-\mathcal{B}'+\frac{h'}{h}\right)\phi'+\left(\frac{(\omega+qA_t)^2}{e^{2(\mathcal{A}-\mathcal{B})}h^2}-\frac{\lambda K^2}{e^{2(\mathcal{A}-\mathcal{B})}h}-\frac{m^2}{e^{-2\mathcal{B}}h}\right)\phi=0\,.
\label{eq:KG-GR}
\end{equation}
This equation is analytically solvable at $\omega=0$ for massless scalar $m^2=0$. The solution is
\begin{multline}
\phi(r)=(r^2+2Q^2-K^2)^{\tilde{q}/2}\biggl[C_1r^{-1+\nu}{_2F_1}\left(\frac{\tilde{q}+3+\nu}{2},\,\frac{\tilde{q}-1+\nu}{2},\,1+\nu;\,\frac{-r^2}{2Q^2-K^2}\right)\\
+C_2r^{-1-\nu}{_2F_1}\left(\frac{\tilde{q}+3-\nu}{2},\,\frac{\tilde{q}-1-\nu}{2},\,1-\nu;\,\frac{-r^2}{2Q^2-K^2}\right)\biggr],\label{eq:sol-GR}
\end{multline}
where
\begin{align}
\tilde{q} &=\sqrt{\frac{2(Q^2-K^2)}{2Q^2-K^2}}\,q\,,\label{eq:qtil-GR}\\
\nu &=\sqrt{\frac{2Q^2+(\lambda-1)K^2}{2Q^2-K^2}}\,.\label{eq:nu-GR}
\end{align}
This solution enables us to extract the analytic solution for zero modes.

\subsection{Matching the IR and UV data}\label{sec:IR-2}
For the extremal geometry, the horizon $r=0$ is an irregular singularity in~\eqref{eq:KG-GR}. When it is sufficiently close to the extremal horizon, $\omega$-dependent terms cannot be treated as small perturbations no matter how small $\omega$ is. In~\cite{Faulkner:2009wj}, a systematic method is developed for treating extremal black hole systems. We will apply this method to the hyperbolic 2-charge black hole in AdS$_5$.

We divide the geometry into inner and outer regions, as shown in figure~\ref{fig:match} with a different IR geometry. The inner region refers to the IR geometry, in which the Klein-Gordon equation can be exactly solved as~\eqref{eq:phiin-GR} below. The outer region refers to the remaining geometry, in which we can make perturbations for small $\omega$. Then we need to match the inner and outer regions.

The IR geometry is obtained as follows. In the $r\to 0$ limit, the metric becomes
\begin{equation}
ds^2=\left(\frac{r}{Q}\right)^{2/3}\left(-\frac{2Q^2-K^2L^4}{Q^2}\,\frac{r^2}{L^2}dt^2+\frac{Q^2}{2Q^2-K^2L^4}\,\frac{L^2}{r^2}dr^2+\frac{Q^2}{L^2}K^{-2}d\mathbb{H}_3^2\right).
\end{equation}
Therefore, the IR geometry is \textit{conformal to} $\text{AdS}_2\times\mathbb{H}_3$. This can be made more explicit by change of variables
\begin{equation}
\zeta=\frac{L_2^2}{r},\qquad L_2=\frac{QL}{\sqrt{2Q^2-K^2L^4}}\,,
\end{equation}
and the metric becomes
\begin{equation}
ds^2=\left(\frac{L_2^2}{Q\zeta}\right)^{2/3}\left[\frac{L_2^2}{\zeta^2}\left(-dt^2+d\zeta^2\right)+\frac{Q^2}{L^2}K^{-2}d\mathbb{H}_3^2\right].\label{eq:ads2}
\end{equation}
The gauge field $A_t$ becomes
\begin{equation}
A_t=\sqrt{\frac{2(Q^2-K^2L^4)}{2Q^2-K^2L^4}}\,\frac{L_2^3}{Q\zeta^2}\,.
\end{equation}
There is a crucial difference between the RN-AdS black hole and the 2-charge black hole. We will switch back to the $r$ coordinate. In the RN-AdS black hole system, $A_t\sim r$, and thus the electric field $E=\nabla A_t$ is constant at the horizon. In the 2-charge black hole system, $A_t\sim r^2$, and thus the electric field $E=\nabla A_t\sim r$ falls off toward the horizon. Consequently, in the near horizon limit $r\to 0$ ($\zeta\to\infty$), the contribution by the electric field to the Klein-Gordon equation is negligible.

To obtain the IR Green's function, we solve the Klein-Gordon equation in the geometry~\eqref{eq:ads2} without the electric field. The solution of $\phi$ with the infalling boundary condition at the horizon is
\begin{equation}
\phi\sim\zeta {H}_{\nu}^{(1)}(\omega\zeta)\,,\label{eq:phiin-GR}
\end{equation}
where $H_{\nu}^{(1)}(x)$ is the Hankel function of the first kind. By expanding~\eqref{eq:phiin-GR} at $\omega\zeta\to 0$, we obtain
\begin{equation}
\phi=\zeta^{1-\nu}(1+\cdots)+\mathcal{G}(\omega)\zeta^{1+\nu}(1+\cdots)\,.
\end{equation}
The IR Green's function at zero temperature is
\begin{equation}
\mathcal{G}(\omega)=-e^{-i\pi\nu}\,\frac{\Gamma(1-\nu)}
{\Gamma(1+\nu)}\left(\frac{\omega}{2}\right)^{2\nu}.\label{eq:gir}
\end{equation}
The IR scaling exponent $\nu$ given by~\eqref{eq:nu-GR} is always real, in contrast to the case of the hyperbolic RN-AdS black hole, due to a different IR geometry with gauge field. Consequently, there is no IR instability.

In the outer region, the solution at small $\omega$ can be written as
\begin{equation}
\phi(z)=\eta_+(z)+\mathcal{G}_k(\omega)\eta_-(z)\,,
\end{equation}
where
\begin{equation}
\eta_\pm=\eta_\pm^{(0)}+\omega\eta_\pm^{(1)}+\mathcal{O}(\omega^2)\,,
\end{equation}
where $\eta_+^{(0)}$ is normalizable and $\eta_-^{(0)}$ is nonnormalizable at the horizon. At the leading order, the asymptotic behavior near the horizon $r\to 0$ is
\begin{equation}
\eta_\pm^{(0)}\to\zeta^{1\mp\nu}=\left(\frac{2Q^2-K^2}{Q^2}r\right)^{-1\pm\nu},\label{eq:etah-GR}
\end{equation}
which we also use to fix the normalization of the solution. The asymptotic behavior near the AdS boundary $r\to\infty$ is
\begin{equation}
\eta_\pm^{(0)}\to a_\pm^{(0)}+\cdots+\frac{b_\pm^{(0)}+\tilde{b}_\pm^{(0)}\ln r}{r^{4}}+\cdots.\label{eq:etab-GR}
\end{equation}
The Green's function near $\omega=0$ can be obtained by the formula~\eqref{eq:green}.

With the solution~\eqref{eq:sol-GR} at hand, we can obtain analytic solutions for $a_\pm^{(0)}$ and $b_\pm^{(0)}$. The asymptotic behavior of $\phi$ near the horizon $r\to 0$ is
\begin{equation}
\phi\to (2Q^2-K^2)^{\tilde{q}/2}(C_1r^{-1+\nu}+C_2r^{-1-\nu})\,.
\end{equation}
By~\eqref{eq:etah-GR} and~\eqref{eq:etab-GR}, the solutions of $a_\pm^{(0)}$ and $b_\pm^{(0)}$ are
\begin{align}
a_+^{(0)} &=Q^{2-2\nu}(2Q^2-K^2)^{3(-1+\nu)/2}\cdot\frac{\Gamma(1+\nu)}{\Gamma\bigl(\frac{3+\nu+\tilde{q}}{2}\bigr)\Gamma\bigl(\frac{3+\nu-\tilde{q}}{2}\bigr)}\,,\label{eq:a0p-GR}\\[3pt]
b_+^{(0)} &=a_+^{(0)}\biggl\{\frac{(2Q^2-K^2)^2}{64}[(1+\nu)^2-q^2][(1-\nu)^2-q^2]\times\nonumber\\
&\qquad\times\Bigl[-3+4\gamma+2\ln(2Q^2-K^2)+2\psi\Bigl(\frac{3+\nu+\tilde{q}}{2}\Bigr)+2\psi\Bigl(\frac{-1+\nu-\tilde{q}}{2}\Bigr)\Bigr]\biggr\},\\[3pt]
a_-^{(0)} &=\bigl.a_+^{(0)}\bigr|_{\nu\to -\nu}\,,\\[3pt]
b_-^{(0)} &=\bigl.b_+^{(0)}\bigr|_{\nu\to -\nu}\,,
\end{align}
where $\psi$ is the digamma function defined by $\psi(x)=\Gamma'(x)/\Gamma(x)$, and $\gamma$ is Euler’s constant.

The zero mode is determined by $a_+^{(0)}=0$. By~\eqref{eq:a0p-GR}, we obtain an analytic solution for the zero modes at the quantum critical point:
\begin{equation}
\tilde{q}-\nu-3=2n\,,\qquad n=0,1,2,\cdots,
\end{equation}
where $\tilde{q}$ given by~\eqref{eq:qtil-GR} is proportional to the charge of the scalar field, and $\nu$ given by~\eqref{eq:nu-GR} is the IR scaling exponent. The onset of the instability is triggered by the most unstable zero mode, which is at $n=0$.

\section{Discussion}
We have analytically solved the zero modes triggering the instability in two systems of charged hyperbolic black holes at zero temperature. Phase transitions of charged hyperbolic black holes imply phase transitions of charged R\'{e}nyi entropies with the entangling surface being a sphere. The main conclusions are summarized as follows.

\begin{itemize}
\item We obtain an analytic solution of the Klein-Gordon equation at $\omega=0$ for a charged scalar field in the hyperbolic RN-AdS$_5$ black hole background. The condition for zero modes is analytically solved. There are two types of instability: the first type of instability happens when the condition for zero modes is satisfied, and the second type of instability happens when the IR scaling exponent becomes imaginary.

\item When the charge of the scalar field is large, both types of instability are possible. After a closer examination, we conclude that the IR instability always comes before the zero mode pole for the RN-AdS$_5$ black hole.

\item We obtain an analytic solution of the Klein-Gordon equation at $\omega=0$ for a massless charged scalar field in the hyperbolic black hole background in the Gubser-Rocha model. The condition for zero modes is analytically solved. The IR scaling exponent is always real, and thus the zero mode genuinely signals the onset of the instability.

\item The condition for zero modes is non-universal. However, analytic solutions are rare.
\end{itemize}

It would be interesting to construct the endpoint of the instability.

\acknowledgments
I thank L.-Y. Hung for helpful conversations. The author is supported in part by the NSF of China under Grant No. 11905298 and the 100 Talents Program of Sun Yat-sen University under Grant No. 74130-18841203.

\appendix

\section{Fermi surfaces from the Gubser-Rocha-axions model}
\label{sec:FS}

After calculating the correlation function of scalar operators, we can also calculate the correlation function of fermionic operators. While zero modes of the former indicate instability, zero modes of the latter indicate Fermi surfaces in the gravitational dual description of certain strongly interacting fermionic systems at finite charge density \cite{Lee:2008xf,Liu:2009dm,Cubrovic:2009ye}. The Green's function of a fermionic operator can be calculated by solving the Dirac equation for a bulk spinor field $\Psi$  \cite{Iqbal:2009fd}:
\begin{equation}
[\gamma^\mu(\nabla_\mu-iqA_\mu)-m]\Psi=0\,,\label{eq:tilpsi}
\end{equation}
where $q$ is the charge of $\Psi$. We solve the Dirac equation in the background of planar black holes closely related to hyperbolic black holes.

Consider an Einstein-Maxwell-dilaton system with $(d-1)$ axion (massless scalar) fields:
\begin{equation}
S=\int d^{d+1}x\sqrt{-g}\left(R-\frac14 Z(\phi)F^2-\frac12(\partial\phi)^2-V(\phi)-\frac{1}{2}\sum_{i=1}^{d-1}(\partial\chi_i)^2\right),\label{eq:axions}
\end{equation}
where $\chi_i=\alpha x_i$ satisfies the equation of motion of $\chi_i$ with the metric
\begin{equation}
ds^2=e^{2\mathcal{A}(r)}\left(-h(r)dt^2+\sum_{i=1}^{d-1}dx_i^2\right)+\frac{e^{2\mathcal{B}(r)}}{h(r)}dr^2.\label{eq:ABh-ansatz-0}
\end{equation}
This system was used as a simple way to introduce momentum dissipation, since $\chi_i=\alpha x_i$ breaks the translation symmetry \cite{Andrade:2013gsa,Gouteraux:2014hca}.

For an arbitrary potential $V(\phi)$, it was observed in~\cite{{Ren:2019lgw}} that the equations of motion for a planar black hole with axions are the same as the equations of motion for a hyperbolic black hole without axions with the metric
\begin{equation}
ds^2=e^{2\mathcal{A}(r)}\left(-h(r)dt^2+d\tsp\Sigma_{d-1,\kappa}^2\right)+\frac{e^{2\mathcal{B}(r)}}{h(r)}dr^2,\label{eq:ABh-ansatz}
\end{equation}
where
\begin{equation}
d\tsp\Sigma_{d-1,\kappa}^2=\frac{d\bar{r}^2}{1-\kappa\bar{r}^2}+\bar{r}^2d\Omega_{d-2}^2\,,\qquad \kappa=-\frac{1}{2(d-2)}\alpha^2,
\end{equation}
where $d\Omega_{d-2}^2$ is the metric for a $(d-2)$-dimensional unit sphere, and $d\tsp\Sigma_{d-1,\kappa}^2$ is the metric for a hyperbolic space with curvature $\sqrt{|\kappa|}$. This relation between a planar black hole with axions and a hyperbolic black hole was observed earlier in~\cite{Gouteraux:2014hca} for a special class of the potential $V(\phi)$.

The simplest background geometry to have holographic Fermi surfaces is the RN-AdS black hole. No analytic solution for the Fermi momentum is available even in the case without axions. In~\cite{Gubser:2012yb}, an analytic solution for the Fermi momentum was obtained from a dilatonic black hole in AdS$_5$ in the Gubser-Rocha model. We will generalize this result to the case with axions.

To solve the Dirac equation, we write $\Psi=(-gg^{rr})^{-1/4}e^{-i\omega t+ikx}\hat{\Psi}$, where $\hat{\Psi}=(\psi_1,\psi_2)^T$. We will focus on $\psi_1\equiv (u_1,u_2)^T$ in the following. Define $u_\pm=u_1\pm iu_2$. For a choice of gamma matrices in~\cite{Gubser:2012yb}, we have
\begin{align}
u_+'+\bar{\lambda}(r)u_+ &=\bar{f}(r)u_-\,,\label{eq:uplus}\\
u_-'+\lambda(r)u_- &=f(r)u_+\,,\label{eq:uminus}
\end{align}
where
\begin{equation}
\lambda(r)=i\sqrt{\frac{|g^{tt}|}{g^{rr}}}(\omega+qA_t)\,,\qquad
f(r)=\frac{m}{\sqrt{g^{rr}}}-ik\sqrt{\frac{g^{xx}}{g^{rr}}}\,.
\end{equation}

When $\omega=0$, the boundary condition for at the horizon is that the solution is regular. The solution for $u_\pm$ with $m=0$ can be written as\footnote{The convention of the branch cut implies $(-1)^\alpha:=(-1+i\epsilon)^\alpha=e^{i\pi\alpha}$ and $(-1-i\epsilon)^\alpha=e^{-i\pi\alpha}$.}
\begin{equation}
u_-=\left(\frac{r}{r+i\sqrt{2}\tilde{Q}}\right)^{\nu}\left(\frac{r+i\sqrt{2}\tilde{Q}}{r-i\sqrt{2}\tilde{Q}}\right)^{\tilde{q}/2}
{_2F_1}\left(\nu-\tilde{q}+\frac{1}{2},\,\nu;\,2\nu+1;\,\frac{2r}{r+i\sqrt{2}\tilde{Q}}\right)\label{eq:umsol}
\end{equation}
and
\begin{equation}
u_+=(-1)^{-\nu+\tilde{q}+1/2}u_-^*\,,\label{eq:upsol}
\end{equation}
where
\begin{align}
\tilde{q} &=\sqrt{\frac{2(Q^2-K^2)}{2Q^2-K^2}}\,q\,,\qquad \tilde{Q}=\sqrt{Q^2-K^2/2}\,,\\
\nu &=\frac{k}{\sqrt{2}\tilde{Q}}\,.
\end{align}
The chemical potential $\sqrt{2}\tilde{Q}$ is a unit of the energy scale.

By perturbation, we can obtain the analytic solution of the Green's function near the Fermi surface. The Green's function can be written as
\begin{equation}
G_R(\omega,k)=\frac{Z}{-\omega+v_F(k-k_F)-\Sigma(\omega,k_F)}\,.\label{eq:green-fm}
\end{equation}
The Fermi momenta are determined by
\begin{equation}
\frac{k_F^{(n)}}{\sqrt{2}\tilde{Q}}=\tilde{q}-n-\frac{1}{2}\,,
\end{equation}
where $n=0$, $1$, $2$, $\cdots$, $\lfloor \tilde{q}-1/2\rfloor$. The corresponding Green's function exhibits one or more Fermi surfaces if $\tilde{q} > 1/2$.

\section{Mathematical notes}
\label{sec:math}
If $\Delta_+$ is an integer, \eqref{eq:sol-RN} is no longer a general solution. If $2\nu$ is not an integer, the general solution for~\eqref{eq:KG-RN} at $\omega=0$ can be written as
\begin{align}
\phi(z)= &C_1\frac{z^{\Delta_+}(1-z^2)^{-1/2-\nu}}{((2-K^2)z^2+1)^{-1/2-\nu+\Delta_+/2}}\times\nonumber\\
&\qquad {_2F_1}\Bigl(\frac{\Delta_+-1}{2}-\nu-\tilde{q},\,\frac{\Delta_+-1}{2}-\nu+\tilde{q};\,
1-2\nu;\,\frac{1-z^2}{(2-K^2)z^2+1}\Bigr)\nonumber\\
+ &C_2\frac{z^{\Delta_+}(1-z^2)^{-1/2+\nu}}{((2-K^2)z^2+1)^{-1/2+\nu+\Delta_+/2}}\times\nonumber\\
&\qquad {_2F_1}\Bigl(\frac{\Delta_+-1}{2}+\nu-\tilde{q},\,\frac{\Delta_+-1}{2}+\nu+\tilde{q};\,
1+2\nu;\,\frac{1-z^2}{(2-K^2)z^2+1}\Bigr).\label{eq:sol2-RN}
\end{align}

\textit{Whittaker function.} Whittaker's equation is
\begin{equation}
\frac{d^2W}{dz^2}+\left(-\frac{1}{4}+\frac{\lambda}{z}+\frac{1/4-\mu^2}{z^2}\right)W=0\,.
\end{equation}
We can write the general solution as
$C_1W_{\lambda,\mu}(z)+C_2W_{-\lambda,\mu}(-z)$, where for large $|z|$ one has
\begin{equation}
W_{\lambda,\mu}(z)\sim e^{-z/2}z^\lambda(1+\cdots)\,,\qquad |z|\to\infty\,.
\end{equation}

\end{document}